\documentclass[onecollarge,natbib]{svjour2}
\bibpunct{[}{]}{;}{n}{}{,} 
\smartqed  
\usepackage{graphicx}
\usepackage{epsfig}
\usepackage{graphicx,amsmath}
\usepackage{color}
\newcommand\ba{\begin{eqnarray}}
\newcommand\ea{\end{eqnarray}}

\newcommand{\be}{\begin{equation}}
\newcommand{\ee}{\end{equation}}

\newcommand{\bas}{\begin{eqnarray*}}
\newcommand{\eas}{\end{eqnarray*}}

\newcommand{\cc}{\c c}

\newcommand{\bno}{\begin{eqnarray*}}
\newcommand{\eno}{\end{eqnarray*}}

\newcommand{\nl}{\newline}

\newcommand{\hsp}{\hspace{0.65cm}}

\journalname{Few-Body Systems}
\begin{document}
\title{
\vspace{-2.0cm} 
\begin{flushright}
{\large \bf LFTC-18-3/24} \vspace{1cm}
 \end{flushright} 
Spin-1 particles and  perturbative QCD
}
\author{J.~P.~B.~C.~de~Melo \and T.~Frederico \and Chueng-Ryong Ji}
\institute{J.~P.~B.~C.~de~Melo  \at
                Laboratorio de F\'isica 
Te\'orica e Computa\cc\~ao Cient\'ifica, 
Universidade Cruzeiro do Sul, 01506-000, S\~ao Paulo, Brazil. 
                \and T.~Frederico \at
Instituto Tecnol\'ogico de Aeron\'autica, 
12228-900, Sa\~o Jos\'e dos Campos, Brazil.
\and 
Chueng-Ryong Ji \at
Department of Physics, Box 8202, North Carolina State University, Raleigh, NC 27695-8202, USA.
}
\date{Version of \today}
\maketitle

\begin{abstract} 
Due to the angular condition in the light-front dynamics (LFD), the extraction of the electromagnetic 
form factors for spin-1 particles can be uniquely determined taking into account implicitly non-valence 
 and/or the zero-mode contributions to the matrix elements of the 
electromagnetic current.  No matter which matrix elements of the electromagnetic current is used to 
extract the electromagnetic form factors, the same unique result is obtained. As physical observables,
the electromagnetic form factors obtained from matrix elements of the current in LFD must be equal 
to those obtained in the instant form 
calculations.  Recently, the Babar collaboration~\cite{Aubert2008} has analyzed the 
reaction~$e^+~+~e^-\rightarrow \rho^+ + \rho^-$ at $\sqrt{s}=10.58~GeV$ to measure the cross section 
as well as the ratios of the helicity amplitudes~$F_{\lambda'\lambda}$. We present our recent analysis 
of the  Babar data for the rho meson considering the angular condition in LFD to put a stringent test
on the onset of asymptotic perturbative QCD and predict the energy regime where the subleading contributions
are still considerable. 

\end{abstract}

\section{Introduction}
\label{intro}

\hsp
The quantum field theory of strong interactions is Quantum Cromodynamics (QCD)
and the main purpose of this theory is to explain the bound state hadronic systems 
in terms of quarks and gluons, the degrees of freedom of QCD. However, this
task is not so easily fulfilled~\cite{Muta1987}. 
The light-front dynamics is one useful approach to describe hadronic bound states, 
mesons or baryons, in particular, because  of the trivial kinematic 
boost properties~\cite{Terentev1976,Kondratyuk1979,Brodsky1998}. 
Nevertheless, the light-front description with a truncated Fock space breaks the rotational symmetry
as the rotation operator in LFD is dynamical~\cite{Melo1997,Melo1998,Naus1998,Naus1999}.
The main consequence from the violation of the rotational symmetry appears as 
the loss of the covariance of calculated matrix elements (i.e. the helicity amplitudes) 
of the electromagnetic (EM) current~\cite{Melo1997,Melo1998,Choi2004}. 
To restore the covariance of the electromagnetic current matrix elements 
in any effective model calculation of LFD, it is crucial to take into account  
the non-valence contribution as well as the zero-mode contribution. 
Once covariance is restored, then the observables calculated with these matrix elements 
will have the same results as in an instant form basis 
calculation~\cite{Melo1997,Melo1998,Naus1998,Choi1998,Bakker2002}. 

In the case of spin-1 particles, the LFD satisfying both parity and time-reversal invariance may yield 
four matrix elements for the electromagnetic current for the plus component of the electromagnetic current,~$J^+$,~
while only three form factors, i.e, charge, magnetic and quadrupole 
form factors exist. However, the requirement of the angular momentum conservation provides   
the angular condition in LFD and due to the angular condition the number of independent matrix elements
in LFD gets identical to the number of independent physical observables or form factors. 
While we have many possibilities to combine the matrix elements of the electromagnetic current in the Drell-Yan frame, 
the extraction of the physical electromagnetic form factors is 
unique due to the angular condition~\cite{Inna84,Chung1988,Frankfurt1993,Hiller1992,Karmanov1996}. 
Because of the freedom in choosing different matrix elements or the helicity amplitudes, 
one should be very careful in taking into account the non-valence contribution and/or 
zero-modes for the correct extraction of the  form factors 
from matrix elements of the electromagnetic current~\cite{Melo1997,Choi2004, Salme1995}.

Unlike the case of the pseudoscalar mesons (e.g. the pion),  vector mesons (in particular, the $\rho$-meson
that we discuss here) does not have much experimental data in order to compare with  models or theory. 
Some time ago, however, the Babar collaboration has extracted some helicity amplitudes for the
reaction~$e^+ e^- \rightarrow  ~\rho^+ \rho^-$~\cite{Aubert2008} and also the cross-section at $\sqrt{s}=~10.58~GeV$. 
In the present work, those experimental data are analyzed to get some insight on the onset of perturbative QCD~(pQCD). 

In the next section, Sect. 2, we briefly discuss the universal ratio for spin-1 electromagnetic form factors.
In Sect. 3, the form factors are related to the helicity amplitudes. In Sect. 4, the sub-leading pQCD contributions are
constrained by the light-front angular condition. In Sect. 5, the available BaBar data are analyzed by taking into account
the sub-leading pQCD contributions. Summary and conclusion follows in Sect. 6.

\section{Universal ratios and spin-1 EM form factors}

In the reference~\cite{Hiller1992}, the authors eliminate the helicity amplitude~$I^+_{11}$ in the light-front spin basis,
 and obtain 
the following expressions for the rho meson electromagnetic form factors (FFs), where the charge, magnetic and quadrupole
form factors  are given respectively by:
\begin{eqnarray}
G_C^{(BH)}&=&\frac{1}{3 (2p^+) (1+ 2 \eta)} \left[ (3-2 \eta) I^{+}_{00}+
8 \sqrt{2 \eta} I^{+}_{10}+2 (2 \eta -1) I^{+}_{1-1} \right],
\nonumber \\
G_M^{(BH)}&=&\frac{2}{ 2 p^+(1+2 \eta)} \left[ I^{+}_{00}
-I^{+}_{1-1}+\frac{(2 \eta -1)}{\sqrt{2 \eta}} I^{+}_{10} \right],
\nonumber \\
G_Q^{(BH)}&=&\frac{1}{ (2 p^+)  (1+ 2 \eta)} \left[ \frac{2}{\sqrt{2 \eta}} I^{+}_{10}
- I^{+}_{00} -( \frac{\eta+1}{\eta}) I^{+}_{1-1} \right]~, \label{BH}
 \end{eqnarray}
 here $\eta=q^2/4 m^2_\rho$.
 
Within  the pQCD, the helicity conservation dominates at very large momentum.  From Eq.(\ref{BH}), the predicted 
universal ratios for the charge form factor, $G_C$, magnetic form factor, $G_M$, and quadrupole
form factor, $G_Q$, are obtained as
\begin{equation}
\label{ratio0}
 G_C:G_M:G_Q~=~\left(1 -\frac{2}{3} \eta   
 \right):2:-1 \, ,
\end{equation}
and the asymptotic relation~\cite{Carlson1984} given by
\begin{equation}
\label{AsymptoticRelation}
 G_c \approx \frac{2 \eta }{3} G_Q
\end{equation}
is consistent with perturbative relations for the EM FFs, at high momentum 
transfers, i.e, $\eta >>1$. 

In the present work, using  the experimental amplitudes from the 
reaction~$e^- e^+  \rightarrow \rho^+ \rho^-$, measured by Babar collaboration~\cite{Aubert2008}, at 
$\sqrt{s}=\sqrt{q^2} =10.58~GeV$,  we study the onset of the perturbative regime 
of QCD. The Babar data amplitudes are given in the time-like regime. 
Nevertheless, as stated by the Phragm\'en-Lindel\"{o}ff theorem~\cite{Denig2013} for analytic functions, space-like (SL) and time-like (TL) amplitudes are the same at very high momentum transfers. As a consequence, the 
EM  FFs are the same for both regimes. 
A previous analysis for Babar data was done in Ref.~\cite{Dbeyssi2012}. However, 
their parametrization of the data was not fully consistent with pQCD, and 
they concluded that their analysis
does not take care of other reaction mechanisms
that could contribute to the $\rho$-meson production in order to satisfy the helicity conservation
(see this discussion in \cite{Melo2016}).

In the case of spin-1 particles, the electromagnetic form factors
could be calculated with the plus component of the current within the 
light-front approach.  In this case, the 
angular condition must be satisfied, which relates
the matrix elements of the electromagnetic current used in Ref.~\cite{Hiller1992}
as well as the other prescriptions given in references~\cite{Inna84,Chung1988,Frankfurt1993}, such that
 the electromagnetic form factors for spin-1  particles are uniquely extracted from the helicity amplitudes 
 in the Drell-Yan frame.

Due to the freedom to chose three out of four matrix elements of the current,
as long as the angular condition is satisfied, we have 
analyzed not only the contribution from $I^+_{00}$~helicity amplitude but also 
the contributions from $I^+_{1-1},I^+_{11},I^+_{10}$~helicity amplitudes to the FFs. 
In order to generalize the universal ratios
and still satisfy the asymptotic relation given by Eq.(\ref{AsymptoticRelation}),
we have now the following expressions for spin-1 ratios of EM FFs, 
\begin{equation}
  G_C:G_M:G_Q~=~\left( \alpha - \frac{2}{3} \eta \right): \beta:-1~.
 \label{ratios2}
 \end{equation}
The values for~$\alpha$,~and~$\beta$ in Eq.~(\ref{ratios2}) was found 
by analyzing the Babar experimental data~\cite{Aubert2008}. 
The present analysis is based on completely model independent first principles:
(i) the light-front angular condition is valid for any~$Q^2$;
(ii) the perturbative power counting rules are valid for~$Q^2 \gg \Lambda^2_{QCD}$; 
and,~(iii) the analyticity of the FFs is valid for entire space-like and time-like regions.

\section{Form factors and helicity amplitudes}

With  the notation  and definitions given in Ref.~\cite{Dbeyssi2012}, 
the EM current of the 
spin-one particle is  written in terms of three Lorentz invariant 
FFs, 
in order to satisfy the covariance, current and parity
conservation. In the TL region, the macroscopic 
EM current is given by~\cite{AR77}: 
\begin{multline}
J_{\mu}=(p_1-p_2)_{\mu}\left[ -G_1(q^2)U_1^*\cdot U_2^*
+\displaystyle\frac{G_3(q^2)}{m_\rho^2}
(U_1^*\cdot q U_2^*\cdot q-\displaystyle\frac{q^2}{2}U_1^*\cdot U_2^*)\right ]+
 \\
+G_2(q^2)(U_{1\mu}^*U_2^*\cdot q-
U_{2\mu}^*U_1^*\cdot q)\,\,,
\label{eq:eq5a}
\end{multline}
where the FFs $G_i(q^2)$ $(i=1, 2, 3)$ are in general complex functions for 
TL momentum transfers ($q^2>0$) and the polarization four-vectors of the
$\rho$ mesons in the final state are $U_1^\mu$ and $ U_2^\mu$.
The electromagnetic FFs are given by 
\begin{equation}
\label{eq:eq6} 
G_C=-\displaystyle\frac{2}{3}\tau (G_2-G_3)+ \left (1-\displaystyle\frac{2}{3}\tau 
\right )G_1, \,\, \,\,\,\,G_M=-G_2,  \,\,\,\,\,\, G_Q=G_1+G_2+2G_3\, ,
\end{equation}
where $\tau={q^2}/{4m_\rho^2}$. 

The helicity amplitudes,~$F_{\lambda_1 \lambda_2}$,~(with Jacob-Wick definitions), 
for the reaction~$ \gamma^* \rightarrow \rho^- \rho^+ $, are given by
\begin{equation}
\label{ha}
F_{\lambda_1 \lambda_2}=M_{\lambda_1 \lambda_2}^{\lambda}=
M(\epsilon\to\epsilon^{(\lambda_{\gamma^*})}, U_1\to U_1^{(\lambda_1)}, U_2\to
U_2^{(\lambda_2)}). 
\end{equation}
The helicity states for the vector mesons, are, $\lambda_1=\lambda_{\rho^+}$, 
and,~$\lambda_2= \lambda_{\rho^-}$. The virtual photon have the helicity,
~$\lambda=\lambda_{\gamma^*}$. Helicity conservation implies that
$\lambda=\lambda_1-\lambda_2$ which leads to
$F_{1-1}=F_{-11}=0$, because the virtual photon has spin 1.  
From symmetry properties, we have that ~$F_{-1-1}=F_{11}$, and, 
also,~$F_{10}=-F_{01}=F_{-10}=-F_{0-1}$, and thus only 
three independent helicities amplitudes remain. 

In Ref.~\cite{Dbeyssi2012}, instead of using the helicity amplitudes 
defined above, the authors used the Breit helicity amplitudes,
$F^B_{\lambda_1 \lambda_2}$,
which are given by
\begin{equation}
F^B_{00} = F_{00} - 2 \tau F_{11}, \,\,\, F^{B}_{10}  =  F_{10}, \,\,\,
 F^B_{11}    =  F_{11}.
\end{equation}
Using the above definitions, the Breit helicity amplitude in terms of the EM FFs are written as:
\begin{eqnarray}
&& F^B_{00}=2\,m_\rho\sqrt{\tau-1}\,\left[G_C-\frac{4}{3}\,\tau \,G_Q\right], \,\,\, 
F^B_{11}=2\,m_\rho\sqrt{\tau-1}\, \left[G_C+\frac{2}{3}\,\tau \,G_Q\right],\nonumber \\
&& F^B_{10}=2\,m_\rho\sqrt{\tau\, (\tau-1)}\, G_M~.
\label{eq:eqHA1}
\end{eqnarray}
Trivially, the  EM FFs can be also written in terms of the 
light-front helicity basis amplitudes.

\section{Angular condition and sub-leading contributions}

The present  analysis of sub-leading order contributions to the 
helicity matrix elements of the EM current starts, for 
convenience, in SL region. The angular condition for the matrix elements
of the plus component of the EM current  in the Drell-Yan frame, and in 
light-front spin basis is given by 
\begin{equation}
 (1+ 2 \eta) I^+_{11} + I^+_{1-1} - 2 \sqrt{2 \eta } I^+_{10} - I^+_{00}=0~. 
\end{equation}
In Ref.~\cite{Inna84}, the angular condition equation was used 
to eliminate the matrix element $I^+_{00}$, and  the following expressions 
for the EM FFs were obtained:
\begin{eqnarray}
\label{gs}
G_C & = &  \frac{1}{2 p^+} \left[  
\frac{3- 2 \eta }{3}  I^+_{11} + \frac{4 \eta }{3} \frac{I^+_{10}}{\sqrt{2 \eta }} + 
\frac{1}{3} I^+_{1-1}
\right] ,\,\,\,
G_M  =   \frac{2}{2 p^+} \left[ I^+_{11} - \frac{1}{\sqrt{2 \eta}} 
I^+_{10}
\right],
\nonumber \\
G_Q & = &  \frac{1}{2 p^+} \left[  
-I^+_{11} + 2\frac{1}{ \sqrt{2\eta} } I^+_{10} -   \frac{I^+_{1-1}}{\eta}
\right]~.
\end{eqnarray}
The above expressions do not contain explicitly the 
matrix element $I^+_{00}$, used to derive the universal ratios 
for the EM FFs in  Ref.~\cite{Hiller1992}. Therefore, in order to bring consistency between
 (\ref{gs}) and the Brodsky-Hiller prescription (\ref{BH}) at large momentum transfers, we have to   
consider pQCD sub-leading contributions such that the angular condition is satisfied. We found the following
relations from the angular condition constraint:
\begin{equation} \label{I+}
\frac{  I^+_{10}}{I^+_{00}}  =   \frac{c_1}{\sqrt{\eta}} + \frac{c_2}{\eta \sqrt{\eta}},
  \,
 \frac{  I^+_{11}}{I^+_{00}} 
  =  \frac{\sqrt{2}c_1 + \frac{1}{2} }{\eta} \,
  , 
\,
\frac{  I^+_{1-1}}{I^+_{00}}   =  \frac{ 2 \sqrt{2} c_2 -(\sqrt{2} c_1 + \frac{1}{2})  }
{\eta} \, ,
 \end{equation}
which is analytically extended from the SL to the TL region, 
keeping the parameters $c_1$ and $c_2$ for both regions.
The new parametrization of the SL FFs taking into account sub-leading contributions follows from the above relations 
and Eqs. (\ref{gs}):
 \begin{eqnarray}
 \label{GSL}
 G_C  &=&    \left[ 1+2\sqrt{2}c_1+4\sqrt{2}c_2 -\eta 
 \right] \frac{f^{+}_{00}}{ 6  }, \,\,\,
 G_M =   \left[ 
 1+\sqrt{2} c_1 - \frac{\sqrt{2}}{\eta} c_2  \right]\frac{ f^{+}_{00}}{2  } \, ,\nonumber \\
G_Q  &=&    \left[ -1 + \frac{1 + 2 \sqrt{2}(c_1-c_2) }{\eta}
\right]\frac{f^{+}_{00}}{ 4 } \, ,
 \end{eqnarray}
where $f^+_{00}=I^+_{00}/(p^+ \eta)$. 
With the generalized expression for the asymptotic ratios FFs, we have
\begin{equation}\label{ab}
\alpha=\frac23\left(1+ 2\sqrt{2}c_1+4\sqrt{2}c_2\right)\, \,\,\,\text{and} \,\,\, 
\beta=2 (1+ \sqrt{2} c_1)~.
\end{equation} 
Also, writing $c_1$ and $c_2$  in terms of $(\alpha,\beta)$, one gets
\begin{equation}
\label{c1c2}
c_1=\frac{\beta-2}{2 \sqrt{2}}\,\,\, 
\text{ and } \,\,\,
c_2=\frac{1  + \frac32 \alpha  - \beta}{4 \sqrt{2}}~,
\end{equation}
and then the electromagnetic FFs, in terms of $\alpha$ and 
$\beta$, are given by:
 \begin{multline}\label{gslab}
G_C= \left[\alpha-\frac23 \eta\right]\frac{f^+_{00}}{4}\, , \, \,G_M=\left[\beta+ 
\frac{1}{2\eta}(\beta-3\alpha-2)\right]\frac{f^+_{00}}{4} \, , \,\,
G_Q=-\left[1+\frac{3}{2\eta}(1+\alpha-\beta)\right]\frac{f^+_{00}}{4}~.
\end{multline}
It is easy to get the general asymptotic relations given in Eq.(\ref{ratios2}). 
We observe that the sub-leading contributions to the FFs in Ref.~\cite{KobPAN95} 
are of the same order as the ones proposed in the present work.

\section{BaBar data analysis with sub-leading order}

The SL region FFs, $G_C$, $G_M$, and $G_Q$, are extended analytically 
to the TL region, i.e, 
$\eta \rightarrow -\tau$, 
where $\eta=\frac{Q^2}{4 m^2_{\rho}}, \tau=\frac{q^2}{4 m^2_{\rho}}$, and $Q^2=-q^2$.
The TL formulas are given below, 
\begin{multline}
\label{GTL}
G_C =  \left[\alpha+\frac23 \tau\right]\frac{f^+_{00}}{4}, \,
G_M =  \left[\beta+ 
-\frac{1}{2\tau}(\beta-3\alpha-2)\right]\frac{f^+_{00}}{4}, \,
G_Q =  -\left[1-\frac{3}{2\tau}(1+\alpha-\beta)\right]\frac{f^+_{00}}{4}\, ,
\end{multline}
and the Breit helicity amplitudes used in~\cite{Dbeyssi2012}, 
are
\begin{multline}
F^B_{10}=
\frac{m_\rho}{8}\, \sqrt{1 - \tau^{-1}}\, \left[2+3 \alpha + \beta (4 \tau-2))\right]
\,f^+_{00}  
 \,,
\\
F^B_{00}=m_\rho \,\sqrt{\tau-1} \left[\beta+ \tau-1\right]\,f^+_{00}  
 ,\\
F^B_{11}=\frac{m_\rho}{4}\, \sqrt{\tau-1}\,\left[2 + 3 \alpha - 2 \beta\right]\,f^+_{00} 
\, .
\label{ratiofbtl}
\end{multline}
The expressions above for the Breit helicity amplitudes are used to analyse
the Babar experimental data at $\sqrt{s}=10.58~GeV$~\cite{Aubert2008}:
\begin{small}
\begin{equation}\label{exp}
|F^B_{00}|^2\,:\,|F^B_{10}|^2\,:\,|F^B_{11}|^2 =0.51\pm 0.14\pm0.07\,:\,0.10\pm0.04\pm0.01    
\,:\, 0.04\pm0.03\pm0.01,
 \end{equation}
 \end{small}
given with the following normalization:
\begin{equation}\label{normalization}
|F^B_{00}|^2 + 4 |F^B_{10}|^2 + 2 |F^B_{11}|^2 = 1.
\end{equation}

If we chose the values from the "universal ratios" of the EM FFs, 
namely, $\alpha=1$, and $\beta=2$, we have  the following relation:
\begin{equation}
 |F^B_{00}|^2\,:\,|F^B_{10}|^2\,:\,|F^B_{11}|^2= 
\left(\,4  + 4\,\tau\right)^2\,:\,
\left(\frac{1}{2\sqrt{\tau}}+4\sqrt{\tau}\right)^2\,:\,1~.
\label{ratiofbtl1}
\end{equation}
Using the above relation together with the experimental 
value of $|F^B_{00}|^2=0.51$ and the normalization (\ref{normalization}), we find the following 
relation among the Breit frame helicity amplitudes: 
\begin{equation}
|F^B_{00}|^2\,:\,|F^B_{10}|^2\,:\,|F^B_{11}|^2 =0.51\,:\,1.1\times 10^{-2}  
\,:\, 1.4 \times 10^{-5},
 \end{equation}
in evident disagreement with the Babar experimental data (\ref{exp}), 
which suggests that the asymptotic region is not yet reached at $\sqrt{s}=10.58$~GeV. 
The values for $\alpha$ and $\beta$ with the quoted errors found by 
fitting the  experimental Babar amplitudes with Eqs. (\ref{ratiofbtl}) considering sub-leading corrections are 
shown in Table 1.
\begin{table}[htb] 
\begin{center}
\caption{Extracted values of $\alpha$ and $\beta$ from the BaBar ratios 
(\ref{exp}) at  $\sqrt{s}=\,10.58$~GeV for $\gamma^*\rightarrow \rho^+ +\rho^- $  using 
the expressions for the helicity amplitudes of Eq.~(\ref{ratiofbtl}), 
with the sub-leading 
contributions. The last line gives the zero of $G_C$ in the SL region. Two sets of
$\{\alpha,\beta\}$ values with $\alpha < 0$, i.e. (III) and (IV), 
have no zero of $G_C$ in the SL region.}
\vspace{0.5 cm}
\begin{tabular}{|c ||c ||c ||c ||c |}
\hline
\hline
Solution  & (I)& (II)& (III)& (IV) \\
\hline
$\alpha$    &  23.1 $\pm$ 8.3 & 10.7 $\pm$ 6.4  & -15.6 $\pm$ 8.3 & -19.2 $\pm$ 6.4  \\
\hline
$\beta$     &  6.4 $\pm$ 2.0  & -5.4 $\pm$ 1.2  &  7.2 $\pm$ 2.0  &  -5.0 $\pm$ 1.2  \\
\hline
$Q_0$[GeV] & 9.1  $\pm$  1.6 &  6.2 $\pm$ 1.9  & -  & -  \\
\hline
\hline
\end{tabular}
\end{center}
\end{table}

The charge EM form factor for spin-1 particles, and in particular for the rho meson, has
a zero at the momentum transfer $Q_0$, and, with sub-leading contributions, 
the zero is given by~$Q_0^2=6 m^2_\rho \alpha$,~with values presented 
in Table 1 (last line), according the values for $\alpha$  and $\beta$ 
found. The results presented in  the table show that the sub-leading 
contributions are extremely relevant for the Babar energy of
$\sqrt{s}=10.58~GeV$.


\begin{center}
\begin{figure}[htb]
\vspace{1.5cm}
\includegraphics[scale=.3,angle=0]{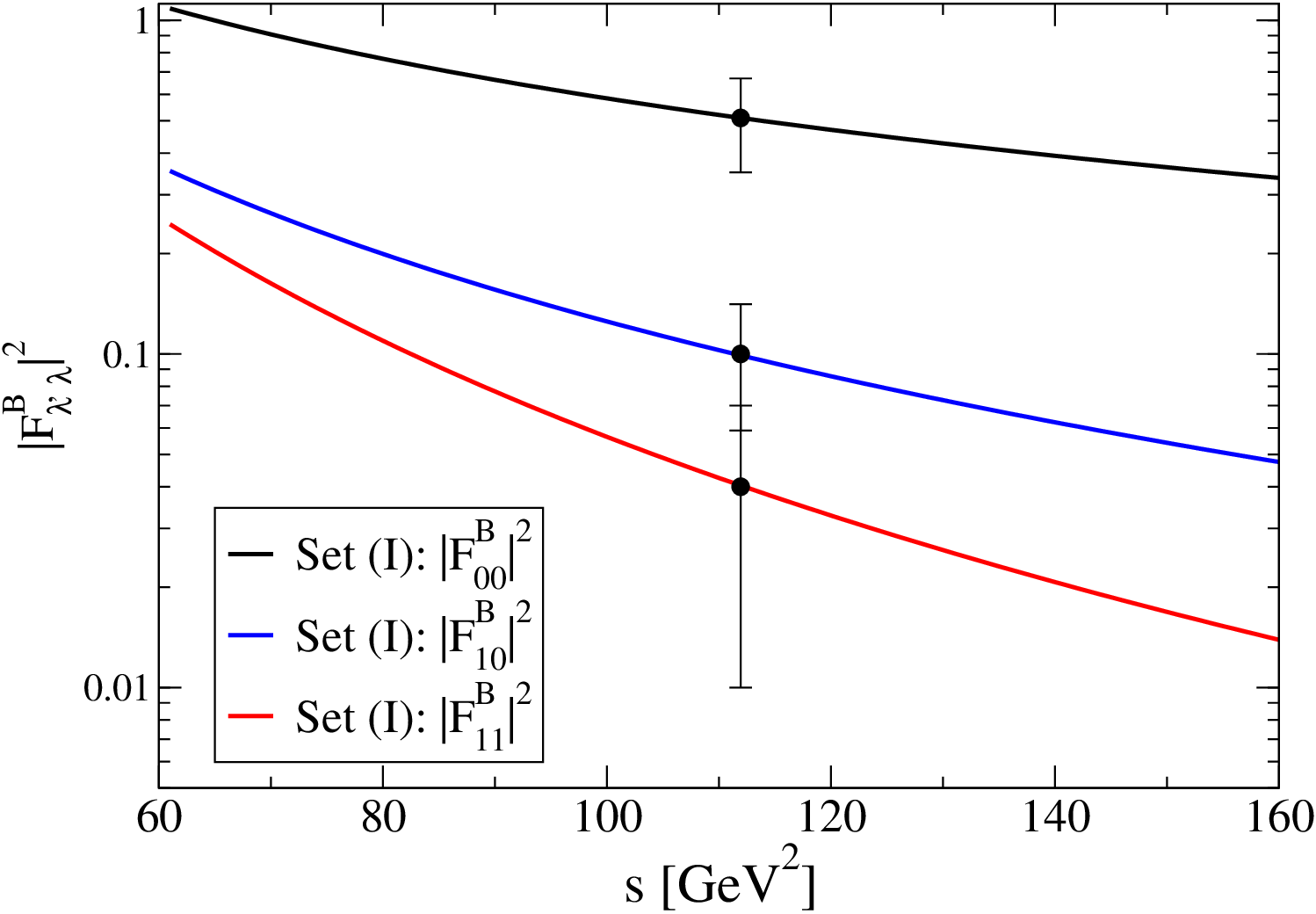}
\hspace{0.1cm}
\includegraphics[scale=.3,angle=0]{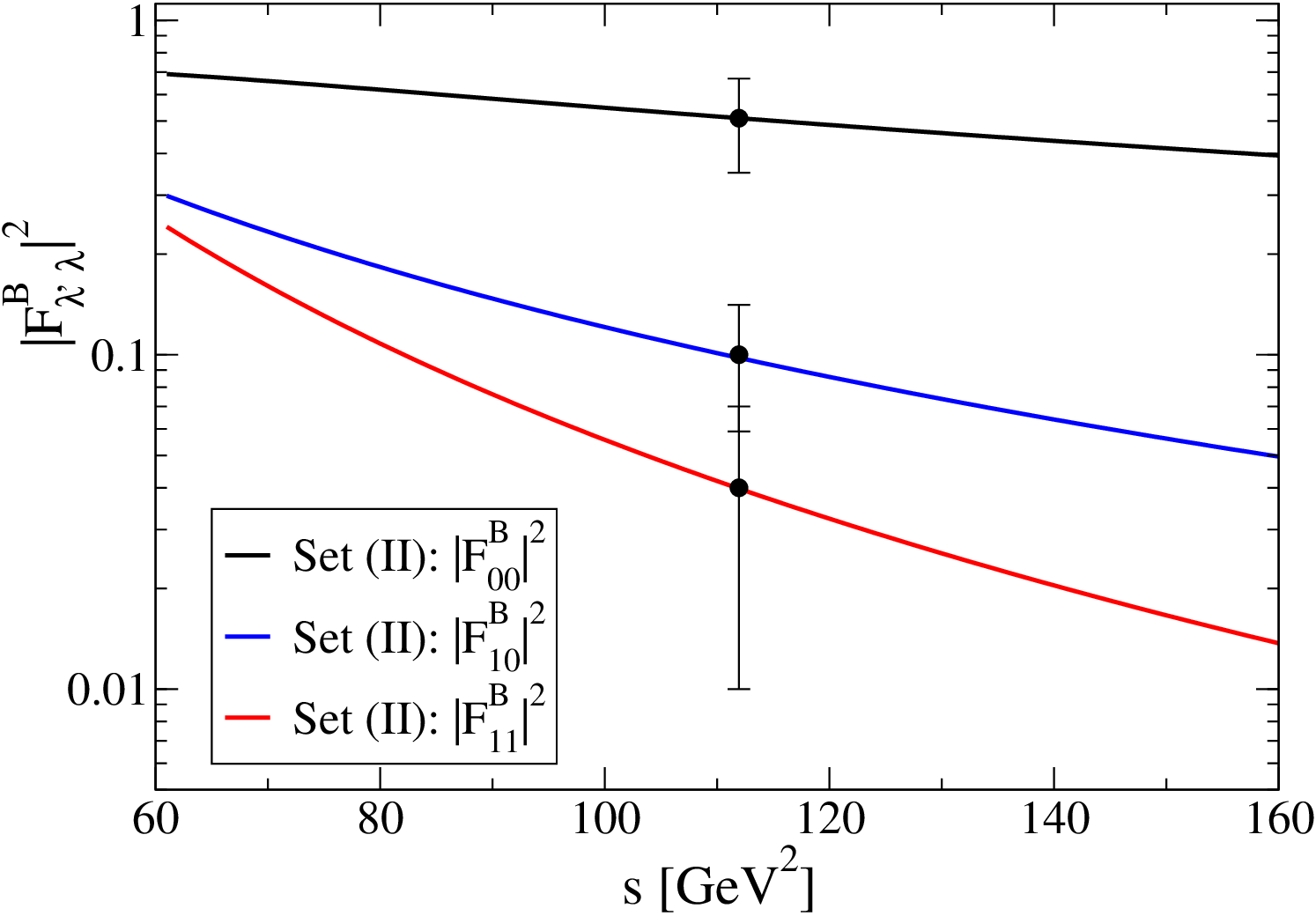}
. \\ \\  \\ 
\includegraphics[scale=0.3]{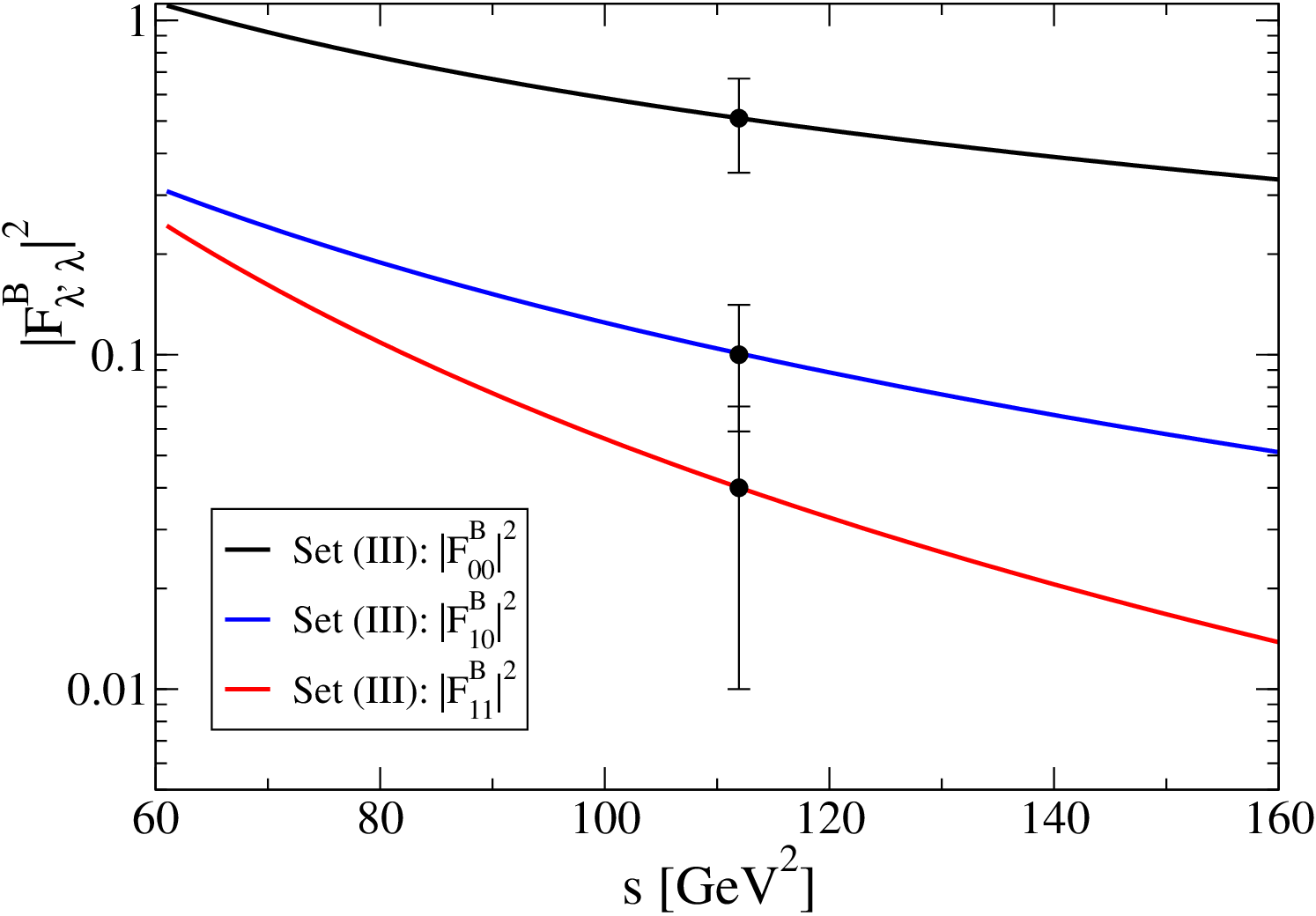}
\hspace{0.10cm}
\includegraphics[scale=0.3]{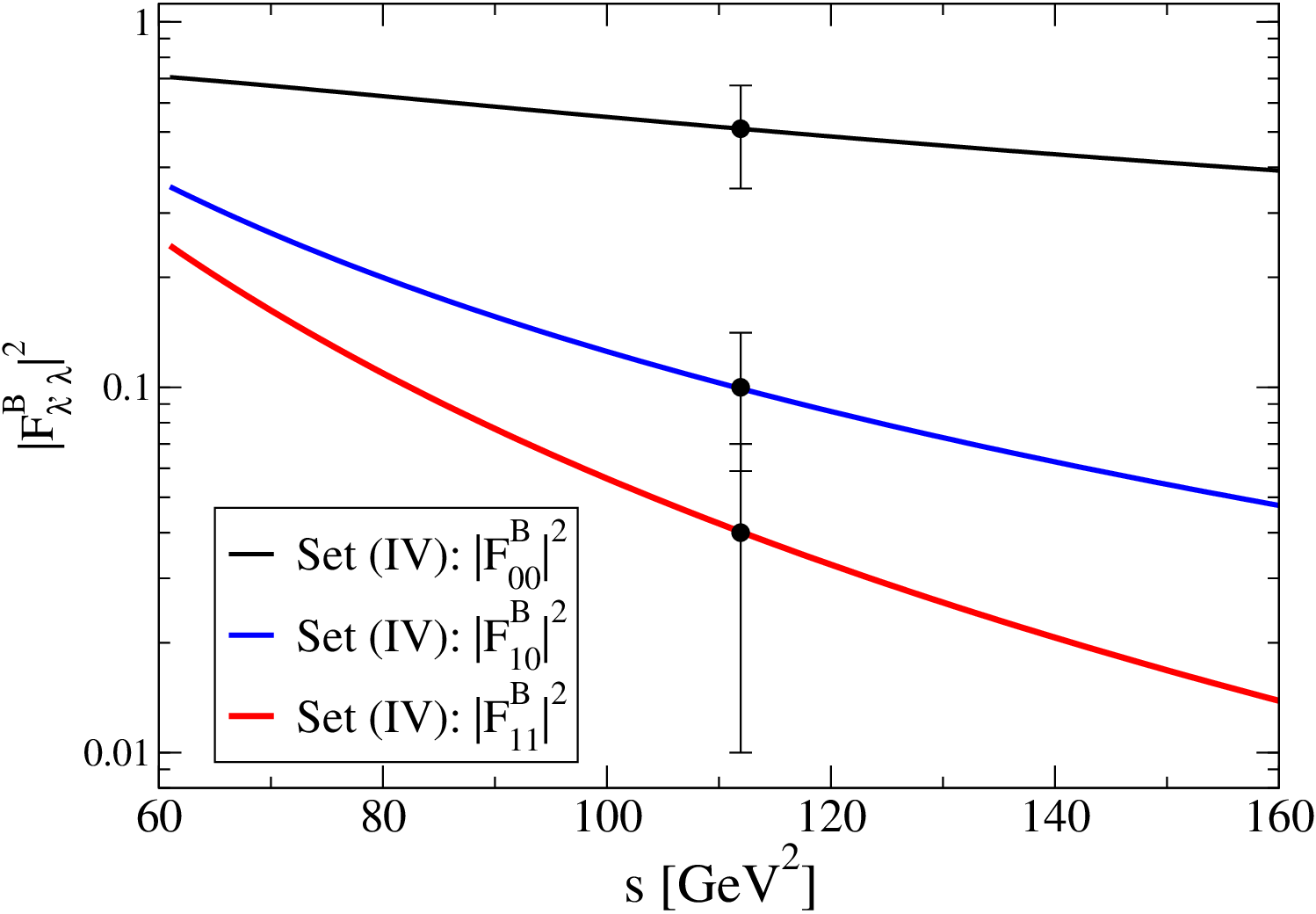}
\caption{(Top-left frame)~$\rho$-meson helicity amplitudes,
$F^B_{\lambda' \lambda}$~for solution (I).
(Top-right frame)~helicity amplitudes for the solution (II). 
The sets (I) and (II) produces a zero for the charge form factor 
in the space-like region. (Bottom-left frame) helicity amplitudes for the solution (III), and,  
(Bottom-right frame) helicity amplitudes for the solution (IV). 
The Babar experimental data 
at~$\sqrt{s}=10.58~GeV$~\cite{Aubert2008}~are also shown in the figure.
}
\label{fig1} 
\end{figure}
\end{center}

\section{Summary and Conclusion}

The four solutions from the fitting of the experimental Babar amplitudes~$F^B_{\lambda'\lambda}$ for 
$\sqrt{s}=10.58$~GeV
with Eq. (\ref{ratiofbtl}) considering sub-leading corrections with the parameters given in Table 1 
are presented in Fig.~1 as a function of the energy square $s$, where also the Babar experimental 
data~\cite{Aubert2008} are seen. The differences in the amplitudes calculated with sets (I)-(IV) are
small in the region of $s$ between  60 and 160~GeV$^2$, but are significant enough to produce the zeros of $G_c$ for solutions 
(I) and (II).  In the figure, we observe that the smooth predicted behaviour of the Breit frame amplitudes 
exhibit a variation of about one order of magnitude in the plotted range of $s$, and eventually could motivate 
future experiments to check the effect of the sub-leading pQCD contributions to the FFs, which were essential to fit
the Babar data. However, the values of $\alpha$ and $\beta$ found are still far from the ``universal" ones, 
clearly indicating that the asymptotic pQCD region is not yet reached at $\sqrt{s}$ = 10.58 GeV. Therefore,
 data at higher energies will be necessary to be used as input to our expressions 
in order to check if $\alpha$ and $\beta$  converge to the ``universal" values of 1 and 2, respectively.
This constitutes an important verification of the pQCD  “universal ratios” of spin-1
form factors.

In summary, we presented expressions for the spin-1 form factor and helicity amplitudes valid at large momentum 
in the SL region that are consistent with the angular condition by  including  sub-leading pQCD contributions.  
The expressions were analytically continued to the TL region and the data for 
the Breit frame helicity amplitudes for the annihilation~(production) 
process $e^+ e^- \rightarrow \rho^+ \rho^-$, from the Babar 
experiment~\cite{Aubert2008} at $\sqrt{s}=10.58$~GeV were analyzed. We hope that our discussion 
can motivate further experimental research on the
asymptotic behavior of spin-1 electromagnetic form factors and the onset of pQCD predictions.

{\it Acknowledgements.} 
This work was partly supported by the Funda\c c\~ao de Amparo \`a Pesquisa do Estado de
 S\~ao Paulo and Conselho Nacional de Desenvolvimento Cient\'ifico e Tecnol\'ogico  (CNPq) of Brazil.
This work is a part of the project INCT-FNA Proc.  No.  464898/2014-5.
This work is also supported in part by the US Department of Energy (No. DE-FG02-03ER41260).

\end{document}